\documentclass[a4paper,num-refs,numbers,sort&compress]{de-rse}

\reviewfrom{23 January 2020}
\reviewto{09 February 2020}
\acceptedon{03 April 2020}
\publicationurl{https://arxiv.org/abs/2005.01469}
\publicationpidtype{arXiv}
\publicationpid{2005.01469 [cs.GL]}

\usepackage{graphicx}
\usepackage{siunitx}

\usepackage[nonumberlist]{glossaries} 
\graphicspath{{figures/}}

\setlist[itemize]{noitemsep, topsep=0pt, leftmargin=*}
\DeclareRobustCommand{\doi}[1]{\href{https://dx.doi.org/#1}{\nolinkurl{doi:#1}}}
\DeclareRobustCommand{\urlhttps}[1]{\href{https://#1}{\nolinkurl{#1}}}
\DeclareRobustCommand{\arxiv}[1]{\href{https://arxiv.org/abs/#1}{\nolinkurl{arXiv:#1}}}
\renewcommand{\glossarysection}[2][]{}

\newcommand*{\vsepfbox}[1]{%
  \begingroup
    \sbox0{\fbox{#1}}%
    \setlength{\fboxrule}{0pt}%
    \mbox{\kern-\fboxsep\fbox{\unhbox0}\kern-\fboxsep}%
  \endgroup
}

\newlist{inlinelist}{enumerate*}{1}
\setlist*[inlinelist,1]{%
  label=(\arabic*),
}

\title{An Environment for Sustainable Research Software in Germany and Beyond: Current State, Open Challenges, and Call for Action}

\author[1,2,\authfn{2}]{Hartwig Anzt} 
\author[1,\authfn{2}]{Felix Bach} 
\author[3,4,5,\authfn{2}]{Stephan Druskat} 
\author[6,7,8,\authfn{2}]{Frank L\"offler} 
\author[9,\authfn{1},\authfn{2}]{Axel Loewe} 
\author[10,11\authfn{2}]{Bernhard Y.\ Renard} 
\author[12,13,\authfn{1}\authfn{2}]{Gunnar Seemann} 
\author[14,\authfn{2}]{Alexander Struck} 
\author[15]{Elke Achhammer}
\author[16]{Piush Aggarwal}
\author[17]{Franziska Appel}
\author[18]{Michael Bader}
\author[19]{Lutz Brusch}
\author[20]{Christian Busse}
\author[21]{Gerasimos Chourdakis}
\author[22]{Piotr W. Dabrowski}
\author[23]{Peter Ebert}
\author[24]{Bernd Flemisch}
\author[25]{Sven Friedl}
\author[26]{Bernadette Fritzsch}
\author[27]{Maximilian D.\ Funk}
\author[3]{Volker Gast}
\author[28]{Florian Goth}
\author[29]{Jean-No\"{e}l Grad}
\author[30]{Sibylle Hermann}
\author[31]{Florian Hohmann}
\author[32]{Stephan Janosch}
\author[33]{Dominik Kutra}
\author[34]{Jan Linxweiler}
\author[10,35]{Thilo Muth}
\author[36]{Wolfgang Peters-Kottig}
\author[37]{Fabian Rack}
\author[38]{Fabian H.C.\ Raters}
\author[39]{Stephan Rave}
\author[40]{Guido Reina}
\author[41]{Malte Reißig}
\author[42,43]{Timo Ropinski}
\author[44]{Joerg Schaarschmidt}
\author[45,46,47]{Heidi Seibold}
\author[48]{Jan P.\ Thiele}
\author[49]{Benjamin Uekerman}
\author[50]{Stefan Unger}
\author[29]{Rudolf Weeber}

\affil[1]{Steinbuch Centre for Computing, Karlsruhe Institute of Technology (KIT), Germany}
\affil[2]{Innovative Computing Lab, University of Tennessee, Knoxville, TN, USA}
\affil[3]{Department of English Studies, Friedrich Schiller University Jena, Germany}
\affil[4]{Institute of Software Technology, German Aerospace Center (DLR), Germany}
\affil[5]{Department of Computer Science, Humboldt-Universit\"at zu Berlin, Germany}
\affil[6]{Heinz Nixdorf Chair for Distributed Information Systems, Friedrich Schiller University Jena, Germany}
\affil[7]{Michael Stifel Center Jena, Germany}
\affil[8]{Center for Computation and Technology, Louisiana State University, Baton Rouge, USA}
\affil[9]{Institute of Biomedical Engineering, Karlsruhe Institute of Technology (KIT), Germany}
\affil[10]{Bioinformatics Unit (MF1), Robert Koch Institute, Berlin, Germany}
\affil[11]{Hasso Plattner Institute, Faculty of Digital Engineering, University of Potsdam, Germany}
\affil[12]{Institute for Experimental Cardiovascular Medicine, University Heart Centre Freiburg Bad Krozingen, Germany}
\affil[13]{Faculty of Medicine, University of Freiburg, Freiburg, Germany}
\affil[14]{Matters of Activity. Image Space Material. Cluster of Excellence at Humboldt-Universität zu Berlin, Germany}
\affil[15]{Technische Universität München, Germany}
\affil[16]{Language Technology Lab, Universität Duisburg-Essen, Germany}
\affil[17]{Leibniz Institute of Agricultural Development in Transition Economies (IAMO), Halle (Saale), Germany}
\affil[18]{Department of Informatics, Technical University of Munich
, Germany}
\affil[19]{Center for Information Services and High Performance Computing (ZIH), Technische Universität Dresden, Germany}
\affil[20]{Deutsches Krebsforschungszentrum, Heidelberg, Germany}
\affil[21]{Chair of Scientific Computing in Computer Science, Technical University Munich, Germany}
\affil[22]{School of Computing, Communication and Business, Hochschule für Technik und Wirtschaft Berlin, Germany}
\affil[23]{Center for Bioinformatics Saar, Saarland Informatics Campus, Germany}
\affil[24]{Institute for Modelling Hydraulic and Environmental Systems, University of Stuttgart, Germany}
\affil[25]{Berlin Institute of Health, Germany}
\affil[26]{Scientific Computing, Alfred Wegener Institute, Helmholtz Center for Polar and Marine Research Bremerhaven, Germany}
\affil[27]{Max-Planck-Gesellschaft e.V.}
\affil[28]{Institut für Theoretische Physik und Astrophysik, Universität Würzburg, Germany}
\affil[29]{Institut für Computerphysik, University of Stuttgart, Germany}
\affil[30]{University Library, University of Stuttgart, Germany}
\affil[31]{Zentrum für Medien-, Kommunikations- und Informationsforschung (ZeMKI), Universität Bremen, Germany}
\affil[32]{Max Planck Institute of Molecular Cell Biology and Genetics, Dresden, Germany}
\affil[33]{European Molecular Biology Laboratory, Heidelberg, Germany}
\affil[34]{Center for Mechanics, Uncertainty and Simulation in Engineering, Technische Universität Braunschweig, Germany}
\affil[35]{eScience Division, Federal Institute for Materials Research and Testing, Berlin, Germany}
\affil[36]{Konrad-Zuse-Zentrum für Informationstechnik Berlin (ZIB), Germany}
\affil[37]{FIZ Karlsruhe - Leibniz Institute for Information Infrastructure, Germany}
\affil[38]{Department of Economics, University of Goettingen, Germany}
\affil[39]{Applied Mathematics, University of Münster, Germany}
\affil[40]{Visualization Research Center, University of Stuttgart, Germany}
\affil[41]{Institute for Advanced Sustainability Studies e.V.}
\affil[42]{Institute of Media Informatics, Ulm University, Germany}
\affil[43]{Department of Science and Technology, Linköping University, Sweden}
\affil[44]{Institute of Nanotechnology, Karlsruhe Institute of Technology (KIT),  Germany}
\affil[45]{LMU Munich, Germany}
\affil[46]{Bielefeld University, Germany}
\affil[47]{Helmholtz Zentrum Munich, Germany}
\affil[48]{Institute of Applied Mathematics, Leibniz University Hannover, Germany}
\affil[49]{Energy Technology, Eindhoven University of Technology, The Netherlands}
\affil[50]{Julius Kühn-Institut (JKI), Federal Research Centre for Cultivated Plants, Quedlinburg, Germany}

\authnote{\authfn{1}axel.loewe@kit.edu, gunnar.seemann@universitaets-herzzentrum.de; \authfn{2}Contributed equally.}

\runningauthor{Anzt, Bach, Druskat, Löffler, Loewe, Renard, Seemann, Struck et al.}

\papernumber{1}

\makenoidxglossaries

\newglossaryentry{RPO}
{
    name = {research performing organizations},
    description = {Research groups, departments, faculties, research institutions (universities, research institutions, cross-institutional research groups, etc.), umbrella organizations, such as Helmholtz-Gemeinschaft Deutscher Forschungszentren, Max-Planck-Gesellschaft zur F\"orderung der Wissenschaften, Leibniz-Gemeinschaft, etc}
}
\newglossaryentry{RFO}
{
    name = {research funding organizations},
    description = {Public research funding bodies but potentially also companies, foundations, associations, etc}
}

\newglossaryentry{GeopoliticalUnits}
{
    name = {geopolitical units},
    description = {Governed public units, ranging from cities and councils, over federal states and countries, up to political unions such as the EU. In the context of this paper, the discussion usually focuses on the larger units (countries and political unions)}
}
\newglossaryentry{RSE}
{
    name = {research software engineers (RSEs)},
    description = {People creating and maintaining research software; this group ranges from research-focused software developers, to software engineers with a focus on research; other definitions include other roles, such as research software managers}
}
\newglossaryentry{DomainResearchers}
{
    name = {domain researchers},
    description = {The people doing the research to advance knowledge in a field}
}

\newglossaryentry{GeneralPublic}
{
    name = {general public},
    description = {Lay people that do not necessarily have specific insight regarding a research domain}
}
\newglossaryentry{ScientificLeaders}
{
    name = {research leaders},
    description = {Heads of research groups, such as professors and other people with staff responsibility}
}
\newglossaryentry{Libraries}
{
    name = {libraries (also registries, indices)},
    description = {Infrastructure units of research bodies such as universities, or independent organizations, which gather research outputs and their structured metadata, and provide indices, search, etc}
}
\newglossaryentry{InfrastructureUnits}
{
    name = {infrastructure units},
    description = {Computing centers of research bodies such as universities and other research centers, as well as high-performance computing facilities}
}
\newglossaryentry{Industry}
{
    name = {industry},
    description = {Companies conducting research or profit from available academic research software which they can directly or indirectly apply to their field}
}
\newglossaryentry{OSDevs}{
    name = {independent (open source) developers},
    description = {Project-external software developers who are not employed by the institution(s) carrying out the project.}
}

\begin{document}

\begin{frontmatter}
\maketitle
\begin{abstract}
Research software has become a central asset in academic research.
It optimizes existing and enables new research methods, implements and embeds research knowledge,
and constitutes an essential research product in itself.
Research software must be sustainable in order to understand, replicate, reproduce, and build upon existing research or conduct new research effectively. In other words, software must be available, discoverable, usable, and adaptable to new needs, both now and in the future.
Research software therefore requires an environment that supports sustainability.

Hence, a change is needed in the way research software development and maintenance are currently
motivated, incentivized, funded, structurally and infrastructurally supported, and legally treated.
Failing to do so will threaten the quality and validity of research.
In this paper, we identify challenges for research software sustainability
in Germany and
beyond, in terms of motivation, selection, research software engineering personnel, funding, infrastructure, and legal aspects.
Besides researchers, we specifically address political and academic decision-makers to increase awareness of the importance and needs of sustainable research software practices. In particular, we recommend strategies and measures to create an environment for sustainable research software,
with the ultimate goal
to ensure that software-driven research is valid, reproducible and sustainable, and that software is recognized as a first class citizen in research.
This paper is the outcome of two workshops run in Germany in 2019, at deRSE19 - the first International Conference of Research Software Engineers in Germany - and a dedicated DFG-supported follow-up workshop in Berlin.
\end{abstract}

\end{frontmatter}


\section{Background}
\label{sec:background}
\noindent\vsepfbox{
  \parbox[]{\dimexpr\columnwidth-3\fboxsep-2\fboxrule\relax}{
    Meet Kim, who is currently a post-grad PhD student in researchonomy at the University of Arcadia (UofA). We will follow Kim’s fictional career in order to understand different aspects of research software sustainability. Note that in Kim's world, many of the changes this paper calls for have already been implemented.
    \\
    {\footnotesize (In our example, Kim is a female person. Of course, research software engineers (RSEs) can be any gender.)}
  }
}

\noindent Computational analysis of large data sets, computer-based simulations, and software technology in general play a central role for virtually all scientific breakthroughs of at least the 21st century. The first image of a black hole may be the most prominent recent example where astrophysical experiments and the collection and processing of data had to be complemented with sophisticated algorithms and software to enable research excellence~\cite{
TheEventHorizonTelescopeCollaboration2019c,Nowogrodzki2019
}. Similarly, it is research software that allows us to get a glimpse of the consequences our actions today have on the climate of tomorrow. However, an implication of computer-based research is that findings and data can only be reproduced, understood, and validated if the software that was used in the research process is sustained and their functionality maintained.

At the same time, sustaining research software, and in particular open research software, comes with a number of challenges. Commercial research software often has revenue flows that can facilitate sustainable software development, maintenance, and documentation as well as the operation of adequate infrastructure. However, a large share of researchers base their research on software that was developed in-house or as a community effort. Many of these software stacks can not be sustained -- often because research software was not a first class deliverable in a research project and hence remained in a prototype state, or because of missing incentives and resources to maintain the software after project funding ended.
Another fundamental difference to industrial software development is that most developers of academic research software (often doctoral students or postdoctoral researchers) never receive training in sustainable software development~\cite{SSIanalysis}. In particular, as they see themselves usually as the primary user of a software product, there are virtually no incentives to invest in sustainability measures such as code documentation or portability.
In combination with the predominance of temporary positions in research, this results in a highly inefficient system where millions of lines of code are generated every year that will not be re-used after the termination of the developer's position.
Part of the problem is the reluctance to accept research software engineering as an academic profession that results in a lack of incentives to produce high-quality software: producing high software quality needs sufficient resources, and although the San Francisco Declaration on Research Assessment (DORA~\cite{DORA}) demands a change in the academic credit system, many institutions base promotion and appointments on traditional metrics like the Hirsch index~\cite{Hirsch2005}.
It is obvious that an extraordinary amount of idealism is required to write sustainable code including documentation and installation routines, as well as running infrastructure and giving support to others when resources can be used more profitably in writing scientific publications based on fragile prototype software~\cite{bangerth2014quo,Prins2015}.

Thus, one main factor for the poor sustainability of research software is the lack of long-term funding for research software engineers (RSEs)~\cite{RSEdef} who take care of the appropriate architecture, organization, implementation, documentation, and community interaction for the software, paired with the implementation of measures towards making the software sustainable during and beyond the development process~\cite{SSIRSE}.

In this paper, we describe the state of the practice and current challenges for research software sustainability, and suggest measures towards improvements that can solve these challenges.
The paper is the result of a community effort, with work undertaken during two workshops and subsequent collaborative work across the larger RSE community in Germany.
It has been initiated during a half-day workshop at the first International Conference for Research Software Engineers in Germany (deRSE19) in Potsdam, Germany on June 5th, 2019~\cite{deRSEconf}, and continued during a dedicated two-day workshop in Berlin, Germany on November 7th and 8th, 2019, which was funded by the German Research Foundation (Deutsche Forschungsgemeinschaft, DFG).
Subsequently, the draft produced during the latter event was opened up for collaborative discussion by the German RSE community through \textit{de-RSE e.V. - Society for Research Software}.

We mainly focus on the situation of research software and RSEs in Germany, where funding bodies increasingly acknowledge the importance and value of sustainable research software and related infrastructures.
The DFG, the largest funding body for fundamental research in Germany, for example, opened a call for sustainable research software development~\cite{DFGcall1} at the end of 2016 and a second call for quality management in research software~\cite{DFGcall2} in June 2019. The first call was oversubscribed by a factor of 10-15, a strong indicator of unmet demand. As another example, the 2019 ``Guidelines for Safeguarding Good Research Practice'' codex of the DFG~\cite{DFGcode} now explicitly lists software side-by-side with other research results and data. The FAIR principles for research data~\cite{Wilkinson2016} provide guidelines for data archiving, but enabling full reproducibility and traceability of research software requires additional steps~\cite{hasselbring2019fair}. In consequence, there are ongoing discussions on whether software should be considered as a specific kind of research data or as a separate entity~\cite{lamprechttowards}.

These positive developments notwithstanding, guidelines and policies for sustainable research software
development in Germany are unfortunately still lacking, and long-term funding
strategies are missing. This all leads to unmet requirements and
unsolved challenges that we want to highlight in this paper by
elaborating on
\begin{inlinelist}
    \item why research software engineering needs to be considered an integral part of academic research;
    \item how to decide which software to sustain; 
    \item who sustains research software;
    \item how software can be funded sustainably;
    \item what infrastructure is needed for sustainable software development; and
    \item legal aspects of research software development in academia.
\end{inlinelist}
While we specifically focus on the research software landscape in
Germany, we are convinced that many of the analyses, findings, and
recommendations may carry beyond. We want to address RSEs who are
experiencing similar challenges and newcomers to the field of research
software development, but first and foremost political and academic decision makers to
raise awareness of the importance of and requirements for sustainable software
development. As a community we work hard on overcoming the challenges of
software development in an academic setting, but we need support -- and reliable funding options and institutional recognition in particular -- for
the sake of better research.

\section{Why Sustainable Research Software in the First Place?}
\label{sec:motivation}

\noindent\vsepfbox{
  \parbox[]{\dimexpr\columnwidth-3\fboxsep-2\fboxrule\relax}{
    After graduation, Kim joins a fixed-term researchonomical research project. For her PhD thesis, she wants to crunch some data. Her colleague recommends learning some Boa, which is an all-purpose programming language often used in researchonomy. Luckily, the UofA runs regular Software Plumbery courses for researchers, including a Boa course. Kim takes the course and gains a solid understanding of the basics of the Hash shell, version control with Tig, and the basics of Boa. She starts writing scripts, which help her a lot with the data processing. Unfortunately, Kim’s scripts are quite slow and actually break after she installs a newer version of Boa. She visits the weekly Code Café organized by her university’s central RSE team. The RSEs not only help her update her scripts but also suggest some changes which speed up the computation by a factor of 25.
    \\
    During the next meeting with her PhD supervisor, Kim presents her collection of scripts. The supervisor encourages Kim to create a Boa library from them, as they will be very useful to other researchonomists. Thankfully, Kim’s project PI had applied for 3 RSE person months in their grant, so the project enlists an RSE from the central team. Over the next three months, Kim and the RSE work together to build the library, document it, test it, license it under the permissive Comanche license, update the TigLab repository to let others contribute, introduce automated builds for every code change via a continuous integration platform, and make the library citable. Finally, they release the first major version of the library, named \emph{hal9k} and publish it through the university library’s software portal, where they get a DOI (Digital Object Identifier) for the version as well as a concept DOI for any future versions of the library. Working with the RSE, Kim has gained a good understanding of some methods in software engineering, and she’s thrilled because this also means she’ll be able to get a job with a local tech company once her fixed-term contract has run out.
    \\
    Kim passes her PhD - of which \emph{hal9k} is an important part - with flying colors, and soon citations to her library start appearing in the researchonomic literature. To Kim’s surprise, she also reads a blog post about a citizen science maker project which has used \emph{hal9k} to process researchonomic data measured in a neighborhood of her hometown. She is invited to give a talk at the local office of Siren, a global tech company, which look to adopt \emph{hal9k}, and pay Kim a generous speaker honorarium. So generous in fact, that Kim can pay a student assistant for a full year from the money.
  }
}
Our credibility as researchers in society hinges on the notion of proper
research conduct, also known as ``good research practice''. The
digitalization of research has introduced complex digital research
outputs, such as software and data sets. Although first
recommendations~\cite{katerbow2018} 
and policies~\cite{https://doi.org/10.2312/os.helmholtz.003} exist, they are far from being widely adopted. It is still somewhat unclear how to translate good research practice into good research
software practice, for example in terms
of validity and reproducibility, but also pertaining to the responsible use of resources. The damage
that failing to do so is causing both to the progress of the research community
and to the credibility of academic research in society is becoming increasingly clear
with the growth of the replication crisis - while the lack of universally
agreed-upon and supported good research software practice is not the main
reason for that crisis, it clearly is a contributing factor.

While it is obvious that software qualifies as a potentially re-usable
digital artifact, the additional benefit of not just reproducing a given
scenario, but transferring software use to new problems, domains,
and/or applications, justifies developing research software with a long-term perspective as \emph{sustainable research software}.

In order to support research, a sustainable software must be correct~\cite{hattonChimeraSoftwareQuality2007,changRetraction2006,matthewsFiveRetractedStructure2007}, validatable, understandable, documented, publicly
released, adequately published (i.e. in persistently identifiable form
as software source code~\cite{smithSoftwareCitationPrinciples2016}, and
potentially in an additional paper which describes the software concept, design decisions, and development rationale),
actively maintained, and \mbox{(re-)usable}~\cite{meraliComputationalScienceError2010,barnesPublishYourComputer2010,tseComputerCodeMore2010}.
We also argue that truly sustainable research software must ideally be published under a Free/Libre Open Source Software (FLOSS) license, and follow an open development model, to (1) enable the validation of research results that have been produced using the software, (2) enable the reproducibility of software-based research, (3) enable improvement and (re-) use of the software to support more and better research, and reduce resources to be spent on software development, (4) reduce legal issues (see section below), (5) meet ethical obligations from public funding, and (6) open research software to the general public, i.e., the stakeholder group with arguably the greatest interest in furthering research knowledge and improving research for the benefit of all.

To make software-based research (and with that almost any research)
reproducible, the used software must continue to exist. Furthermore, it must
continue to be usable, understandable, and return
consistent results (or potential changes to results and bug fixes must be clearly documented) in the
evolving software and hardware environment. Moreover, the software
should support reuse scenarios to avoid duplication of efforts and
unneeded drain of resources. Therefore, if research software is publicly
funded, it should be freely available under a FLOSS license.

Currently, creating and using sustainable research software is not sufficiently incentivized. To evaluate in which area this shortcoming should be addressed, we have identified the following challenges: \label{sec:motivation:challenges}

\begin{itemize}
\item
  \textbf{Lack of benefit for the individual}:
  Currently, the primary motivation for
    sustainable research software is the common benefit, rather than the
    individual benefit. It is  clearly beneficial for the
    research community as a whole to direct resources towards sustainable
    research software, as it enables better and more
    research by freeing funds for domain research rather than
    (repetitive) software development. But the developers are often even at a disadvantage (e.g., they publish fewer papers~\cite{bangerth2014quo,Prins2015}), which
    in turn prevents sustainable research software.
\item
  \textbf{Lack of suitable incentive systems}: Contributions to research that are not
  traditional text-based products (i.e., papers or monographs) are still not sufficiently rewarded, or not rewarded at all, due to the missing implementation of mandatory software citation ~\cite{smithSoftwareCitationPrinciples2016,Hafer2009,Howison2016,Li2017,Li2019,Park2019,Pan2019,Doerr2019,druskatSoftwareDependenciesResearch2019,katzSoftwareCitationImplementation2019}, among other reasons. Interestingly, one third of research software repositories have a lifespan (defined as the time from the first time
any code was uploaded to the last contribution) of less than one day (median: 15 days~\cite{hasselbring2019fair}), indicating that many codes are only made available publicly for the publication in a journal (as increasingly encouraged or required by journals~\cite{Resnik2019139} and associated with higher impact~\cite{codeImpact}) but are not maintained thereafter. 
\item
  \textbf{Lack of awareness}: Research software sustainability (see~\cite{Venters2014,Goble2014,Druskat2016,KatzFundamentalsSoftwareSustainability2018}) and its
  importance is lacking visibility as well as acceptance, and research software
  engineering in its implementation as sustainable software development
  and software maintenance is not sufficiently supported, both in
  Germany and beyond~\cite{SSIRSE,NWOprofile,Casties2019}.
\item
  \textbf{Lack of expertise}: Knowledge about how to create, maintain, and
  support sustainable research software is emerging~\cite{Wilson2014,Stodden2014,Wilson2017} but
  has not yet permeated related activities within organizations - specifically
  teaching, mentoring, and consultancy. This lack of expertise can also lead to divergence between
  software design and community uptake, e.g., if the software fails to meet the needs of the
  target group, or is insufficiently usable. RSEs
  combine sustainable software engineering expertise with experience in one or
  more research domains.
\item
  \textbf{Heterogeneous research community}:
  There are significant differences with respect to how software is developed, published, used, and valued in the different academic disciplines. Additionally, there is even heterogeneity within a community in terms of application and approach.
  This also makes it hard to train researchers for sustainable software development,
  as beyond basic training in computational research such as provided by The Carpentries, advanced courses for research software engineering are not widely available (with the notable exception of the CodeRefinery project~\cite{CodeRefinery}).
  Targeted curricula must be developed and updated regularly, and specialized instructors need to be trained.
\item
  \textbf{Lack of impact measures}: It is unclear how to measure the impact
  of research software with respect to its quality, reusability, and benefit for the research community. This exceeds the
  implementation of research software citation (which is work in progress~\cite{smithSoftwareCitationPrinciples2016,liSoftwareCitationReuse2016,katzSoftwareCitationImplementation2019,druskatSoftwareDependenciesResearch2019}), and pertains to sustainability and policy studies.
\item
  \textbf{Infrastructure issues}: Due to a lack of knowledge about
  how sustainability features impact the application of research software, there is not yet enough evidence
  for whether centralized or decentralized facilities should be favored to
  further research software sustainability~\cite{Kuchinke2016,SuLMaSS,Hexatomic}.
  This in turn leads to a lack of infrastructure as a whole.
\item
  \textbf{Legal issues}: Many obstacles for research software pertain to
  legal issues, such as applicable licensing and compatibility of licenses~\cite{morinQuickGuideSoftware2012}, and decisions about license types.
\item
  \textbf{Funding issues}: Despite some individual initiatives~\cite{katzLookingSoftwareSustainability2015,DFGcall1,DFGcall2,ChanZuckerberg}, funding for the creation, maintenance, and
  support of sustainable research software is still scarce.
\item
  \textbf{Slow adoption of research software engineering as a profession}: Career options for
  research software work are not fully determined, although career paths are emerging in some regions.
  Initially, the RSE initiative in the UK has made progress in this area, and RSE groups have been installed in many institutions. In Germany, the
  US, and the Netherlands, this is still work in progress~\cite{SSIRSE,de-rse,us-rse,nl-rse}. It is also not yet
  determined how to match research software engineering roles in public
  institutions with industry roles (see~\cite{rodriguez-sanchezAcademiaFailureRetain2017}).
\end{itemize}

In summary, the necessary but resource-intensive practice of creating, maintaining, supporting, and funding sustainable
research software is not yet sufficiently incentivized and enabled by
research institutions and funding agencies, nor does it align well with
the publish-or-perish culture that is still prominent in most fields.

Therefore, it is necessary to comprehensively motivate sustainable research software
practice.
In the following, we identify stakeholders of research software (see~\cite{Katz2019,Druskat2018,ye2019open}), and explicate their particular motivations for sustainable research software. Subsequently, we
specify challenges towards satisfying the demands of the individual stakeholders.

\subsection{Stakeholder Motivations for Research Software Sustainability}

While a wide range of stakeholders share interest in sustainable software, we argue that their individual motivation can differ quite significantly:

The \textit{\gls{GeneralPublic}} benefits from research which supports the
common good, in other terms: creates a better world, faster. Taxpayers
have an interest in economical use of their tax money, to which
duplicated or flawed efforts to create research software -- in contrast to software reuse -- is contrary. A subset of this group may be interested in sustainable, i.e., re-usable and understandable, software as part of citizen
science.

\textit{\Gls{DomainResearchers}} benefit from better software to do more, better, and faster research. Sustainable research software supports this through validated functionality (e.g., correct algorithms), the potential for reuse, and general availability. Sustainable software also potentially simplifies building upon previous research results by re-using the involved software to produce additional data or by extending the software's functionality.
In light of recent updates to definitions of good research practice~\cite{DFGcode}, sustainable research software also allows domain researchers to comply with guidelines and best practices. Additionally, using a software that is sustainable enough to
establish itself as a standard tool in a field signifies inclusion in a research
community. Less directly, researchers may benefit from the existence of sustainable standard tools as they yield standard formats, which in
themselves facilitate reuse of research data.

\textit{\Gls{RSE}} have an intrinsic interest in
sustainable research software. They create better software for research, which enables more and better research. RSEs have an inherent interest in developing and working with high quality software, as part of professional ethics as well as good research practice. RSEs build their reputation on high quality software and software citation~\cite{smithSoftwareCitationPrinciples2016,druskatSoftwareDependenciesResearch2019}, which will open up new career paths.
Finally, for RSEs, creating sustainable research
software is part of an attractive, intellectually challenging, and
satisfying work environment.

\textit{\Gls{ScientificLeaders}} as well as \textit{\gls{RPO}} mainly focus on the economic aspects and management of research, i.e., available funds, people, and time employed to optimize research output.
Both need to make sure that their employees continually improve their qualification and generate impact to improve their standing in the various research communities and ensure continued funding.
Overseeing and enabling the creation of sustainable
research software advances their visibility in the field and makes
their research endeavors both more future-proof and more easily
traceable, reproducible, and verifiable and thus more likely to attract additional resources (including human resources). \textit{\Gls{RPO}} can additionally benefit from sustainable research software if it can be reused in other areas, creating synergies between different research disciplines.
These synergies typically free resources that can then be used in areas other than software development and maintenance.
Finally, organizations can gain
highly competitive positions in terms of funding and hiring opportunities, as well as a reputation for being on the cutting edge of research, through early adoption of research software
engineering units, and the implementation of sustainable research software
policy and practice.

\textit{\Gls{RFO}} have inherent interest in -- and directly benefit from -- the existence of sustainable research software as it allows them to direct more resources towards actual research (rather than recreation of software) and increase return on investment.
At the same time, funding organizations can create incentives for sustainable software by imposing policies that reflect the necessity of research software sustainability and creating respective funding opportunities.

\textit{\Gls{GeopoliticalUnits}} have a strategic interest to be independent of other geopolitical units to ensure that research can continue seamlessly regardless of geopolitical developments and ensuing embargoes on information flow.
Reuse of sustainable software additionally frees up funding for uses other than software development.
Well-established, sustainable software systems can also attract researchers and companies in the research and technology sector.

\textit{\Gls{Libraries}} benefit from sustainable research software, as it will undergo a formal publishing process and be properly described in its metadata. Libraries can extend their portfolio beyond text-based research objects and stake claims as organizations harnessing the digitalization of research. In turn, they help to increase visibility and discoverability for research software through their services and advance the competitiveness of their organization or geopolitical unit. In addition, libraries also use research software and would thus benefit directly from a more sustainable research software landscape. Last but not least, by using FLOSS research software, libraries could avoid expensive licenses and often insufficiently adapted commercial software.

\textit{\Gls{InfrastructureUnits}}, such as supercomputing facilities and
university computing centers, benefit from sustainable software as it makes their daily work in terms of software installation and user support easier. Additionally, they can position themselves at the forefront of research by bundling expertise on the creation and maintenance of sustainable research software and installing research software engineering teams.

\textit{\Gls{Industry}} benefits from sustainable research software, as the
process of creating and maintaining research software produces a highly-skilled workforce.
Depending on the employed licensing model, sustainable research software can also be adopted by industry partners to reduce cost in corporate research and development.
Helping to sustain research software may also enable positive outreach for companies across industry and into society.

\textit{\Gls{OSDevs}} can get involved in research software, even if they 
are not employed by a research institution. This can help them get in contact with
other developers in the field and may potentially lead to collaborations or job opportunities in research based on this extended experience.

\section{How to Decide Which Software to Sustain?}
\label{sec:selection}

\subsection{Requirements and Challenges}
The sustained funding of all existing software efforts is not only impossible but would risk to overly splinter the community and eventually become counterproductive to the efficiency of the research community. Therefore, it is important to agree on a list of transparent criteria that qualify a software product for sustained funding.
We recognize that defining research software engineering criteria for software evaluation will also lead to activities aiming at optimizing scores to achieve these criteria. Hence, the criteria have to be designed such that all score-pushing effort truly advances the value of the software. Criteria that can be manipulated without effectively adding value, i.e., wasting resources, should be excluded.
The list of criteria presented in this chapter could be the basis for a structured review process that facilitates an unbiased evaluation of software tools from various fields. Therefore, this list must be general enough to be applied to research software from various research disciplines while also respecting differences between fields (e.g. citation rates between humanities and life sciences). The challenge to do justice to a wide spectrum is e.g. reflected by suggesting criteria comprising different levels~\cite{ChueHong2014}.
One of the major challenges in the endeavor to define a selection scheme for sustainable funding of research software is to organize a fair and transparent review process. We believe that it is important that the review process is conducted by experts, or teams of experts, that have a strong background both on software engineering as well as on the domain-specific aspects, the latter because certain criteria often exist on a spectrum that is most likely shaped by the specific demands of the respective research community.

\noindent\vsepfbox{
  \parbox[]{\dimexpr\columnwidth-3\fboxsep-2\fboxrule\relax}{
        Kim’s PI is happy because Kim writes a longer section on \emph{hal9k} for the final project report and provides a software management plan alongside it, which ticks off a box in the template that the PI had previously worried about. The PI does not want to let Kim go and instead offers her to be co-PI on a follow-up project to test new methods on the data, and integrate them into \emph{hal9k} as well. They are positive that such a project proposal has a good chance to be funded, as they can show impact of their first project via their university's current research information system (CRIS) and through the number of citations of \emph{hal9k} and the publications for which it was used. While they write the proposal, the faculty dean approaches the two to tell them that based on Kim’s work, they will now negotiate about two new RSEs for the central RSE team with the university’s provost for research and plan to consider candidates with a background in researchonomics.
        \\
        When they get the decision letter from the research funding organization, Kim and her co-PI are happy to learn that their new project has won the grant. The reviewers specifically point out the value of extending Kim’s Boa library to include the proposed new methods, as well as the significant reuse potential of \emph{hal9k} for the researchonomic community as a direct effect of its well-engineered architecture and modularity. Additionally, they stress that it was really easy to evaluate the software due to the comprehensive test suite, documentation, and example data. In fact, during the first month of the new project, three other researchonomic research projects approach them to ask whether they can contribute to Kim’s library and offer to fund six months of RSE work for this. Kim uses this money to also parallelize \emph{hal9k} together with the RSEs and works with her university’s computing center to offer it as a standard tool for researchonomic supercomputing.
    }
}
\\ \\
\noindent While an assessment based purely on quantitative metrics would allow for seemingly objective comparisons between programs, the definition of valid and robust quantitative metrics that can be evaluated with reasonable effort is a major challenge. On the other hand, a structured qualitative assessment with scores for groups of criteria can provide a middle ground.
It is clear that both preparing an application for a review against these criteria from the applicant side as well as the evaluation by the reviewers requires significant effort. We believe that the added value significantly outweighs the investment but appropriate resources need to be factored in.
Sustainability of research software should be considered from the beginning for new projects. The criteria listed below, or a subset such as the ``good enough'' practices proposed by Wilson et al.~\cite{Wilson2017}, are valuable throughout the development process (including early phases) for almost all types of research software applications. ``Classical'' research funding schemes should acknowledge the need to follow best practices during the development of new software and allow factoring in appropriate resources to design and implement for sustainability. In this section, we focus on the question which software to support in dedicated sustainability funding schemes. For such sustained funding, only software in application class 2 or 3 as defined by Schlauch et al.~\cite{https://doi.org/10.5281/zenodo.1344612}, i.e., with significant use beyond personal or institutional purposes, would likely be considered.
Excellence as reflected in funded projects, publications, and software adoption, i.e., backing by a community, should be considered during selection. Nevertheless, we believe a good scheme should strike a balance between consolidating the field to few well-established software packages on one side and stimulating innovation and cooperation promoting diversity in terms of more than one monopolistic package on the other side.
Last but not least, there is an inherent conflict between the long-term goals of sustainability funding a software and the necessary reevaluation to monitor the state of the software over time.

\subsection{Selection Criteria}
Several evaluation schemes for research software have been proposed before and led to the formulation of first recommendations~\cite{katerbow2018,https://doi.org/10.2312/os.helmholtz.003}. Gomez-Diaz \& Recio suggested the CDUR scheme based on Citation, Dissemination (including aspects like license, web site, contact point), Use, and Research (output)~\cite{GomezDiaz2019}. Lamprecht et al. rephrased the FAIR data principles~\cite{Wilkinson2016} for research software~\cite{lamprechttowards}. Hasselbring et al. found that the adoption of FAIR principles is different between fields with an emphasis on reuse in computer science as opposed to a reproducibility focus in computational science~\cite{hasselbring2019fair}. Fehr et al. collected a set of best practices for the setup and publication of numerical experiments~\cite{Fehr2016}. Jim\'{e}nez et al. boiled it down to four best practices~\cite{Jimnez2017}: public source code, community registry, license, and governance. Hsu et al.~\cite{Hsu2019} proposed a framework of  seven sustainability influences (outputs modified, code repository used, champion present, workforce stability, support from other organizations, collaboration/partnership, and integration with policy). They found that the various outputs are widely accessible but not necessarily sustained or maintained. Projects with most sustainability influences often became institutionalized and met required needs of the community~\cite{Hsu2019}. In the field of open source software, the CHAOSS (Community Health Analytics Open Source Software) project has developed metrics to evaluate sustainability~\cite{CHAOSS}. One objective of CHAOSS is to automatically generate project health reports based on software that evaluates the metrics, with most of the metrics already covered. The UK Software Sustainability Institute (SSI) suggested both a subjective tutorial-based and a more objective criteria-based software evaluation scheme~\cite{SSIevaluation}, the latter being available as an online form~\cite{SSIsustain}. ROpenSci~\cite{ROpenSciDevGuide} provides software reviews for R developers, which have been very successful in the community. The review criteria of the Journal of Open Source Software (JOSS)~\cite{JOSSReview} focus on the aspects license, documentation, functionality, and tests. This list of essential items should be fulfilled by all research software that wants to be considered not only for publication but also for sustained funding. 

We drew inspiration from all these works and suggest a set of criteria to base reviews for sustainable funding on. This set comprises mandatory, hard criteria that we think have to be fulfilled across domains (highlighted in italics) and additional desirable, soft criteria that can be implemented to different degrees depending on the use case and domain-specific software development requirements. The soft criteria should be evaluated in a structured way by the reviewers with a specific response for each section rather than one running text.
The fact that most of these criteria will be considered in any software management plan (SMP,~\cite{SSImanage}) highlights its importance for sustainable research software.

\subsubsection{Usage and Impact}
Requirements qualifying software for sustained funding are 
\begin{inlinelist}
    \item its \textit{use beyond a single research group}, 
    \item the scientific relevance and validity of the software documented in \textit{at least one peer-reviewed scientific publication}. Ideally a paper also describes the scope, performance, and design of the software.
    \item The use of the software in publications is a measure of impact but quantitative assessment brings about additional challenges~\cite{Li2019}. Therefore, other, potentially domain-specific, impact measures, such as influence on policy and practice as well as use in other software and products should be considered as well to evaluate relevance for academia and society. Considerable attendance at training and networking events can be considered as a proof of use as well.
    \item A \textit{market analysis} needs to show that the software is important to a user base of relevant size and either unique or one of the main players in a field with several existing solutions. Geographical or political aspects can be considered as well, e.g. to support the maintenance of a European solution. A convergence process of (parts of) a research community towards a specific software stack, i.e., documented transition of several research groups to a common software, would be a strong indicator of impact.
    \item As community uptake and benefits are a central goal of sustained software funding, outreach and \textit{appropriate training material} for new users of the software are essential.
\end{inlinelist}

\subsubsection{Software Quality}
As mandatory criteria of software quality that have to be fulfilled, we consider 
\begin{inlinelist}[resume]
    \item the \textit{public availability of the source code} in both a code repository and an archive (for long term availability), developed using
    \item \textit{version control} with meaningful commit messages and linked to an issue tracker (ideally maintained, but at least mirrored on a public platform). 
    \item \textit{Documentation} of the software needs to be publicly available comprising both user documentation (requirements, installation, getting started, user manual, release notes) and developer documentation (with a development guide and API documentation within the code, e.g. using Doxygen)~\cite{Lee2018}.
    \item The \textit{license} under which the software is distributed must be defined. Publicly funded software
should be published under a Free/Libre Open Source Software (FLOSS) license
by default, although exceptions to this might apply (e.g. excluding commercial use).
\item \textit{Dependencies} on libraries and technologies must be defined.
\end{inlinelist}

We acknowledge that some additional criteria have to be evaluated under consideration of the research domain. These comprise 
\begin{inlinelist}[resume]
    \item the availability of examples (comprising input data and reference results), 
    \item mechanisms for extensibility (software modularity) as one aspect of software architecture~\cite{Venters2018} and 
    \item interoperability (APIs / common and open data formats for input and output), 
    \item a test suite (including at least some of the following: unit tests, regression tests, integration tests, end-to-end tests, performance tests; ideally run in an automated fashion in a continuous integration environment),
    \item tagged releases (considering their frequency, and availability for end users in terms of binary packages for major operating systems, or availability via package managers or containers),
    \item no large-scale re-implementations for functionality for which good solutions already exist. Many of these aspects require appropriate infrastructure (see page \pageref{sec:infrastructure}).
\end{inlinelist}

\subsubsection{Maturity}
The research software applying for sustained funding must have already reached a certain level of maturity (typically class 2 or 3 as defined by Schlauch et al.~\cite{https://doi.org/10.5281/zenodo.1344612}). A mandatory requirement is 
\begin{inlinelist}[resume]
    \item a comprehensive and up-to-date \textit{software management plan}~\cite{SSImanage}. The software should 
    \item be maintainable with an appropriate amount of resources as detailed in a sustainability section of the software management plan. The software has 
    \item a well maintained website with a clearly defined \textit{point of contact} and a communication channel to inform users about news regarding the software such as new releases. Besides an active user community, sustainable software requires 
    \item a group of developers (i.e., definitely \textit{more than 1 developer}) documented, e.g. by contributions to the code base or participation in documented, public discussions or issue tracking. Another criterion is 
    \item whether potential contributors are invited to participate in a clearly defined process (e.g., a CONTRIBUTING document). The group of developers should have defined a governance model for their project and easy ways for users to provide input regarding their needs.
\end{inlinelist}

\subsection{Recommendations}
Given the diversity in the software technology landscape, and the domain-specific software development cultures~\cite{johanson2018software}, some of the above-mentioned criteria have to be evaluated against domain-specific requirements. Therefore, we highly recommend to base the selection process on a combination of 
\begin{inlinelist}
    \item a software quality-based review and 
    \item a domain-specific scientific review.     
\end{inlinelist}
In particular, the former should be ideally performed by a central institution (e.g. at funding bodies or other independent agencies such as a software sustainability institute). Only criteria for which improvement truly advances the value of the software should be considered in evaluation schemes, i.e., no criteria that can be gamed.
After rejecting software not fulfilling the mandatory criteria in a first stage of the review process, the second stage of the selection process should be realized as a transparent procedure ideally allowing the reviewers to interact with the PIs of the software (e.g. remote meetings, forum-like discussions) and put the software quality and development efforts into the domain-specific context. The outcome of this second stage should be a structured review assessing each criterion explicitly and a rating for each of the dimensions \emph{Usage and Impact, Software Quality, and Maturity}.
For sustained software funding, it is important to audit the performance, relevance, impact, progress, and level of sustainability of funded software after reasonable time frames. Such a reevaluation should revisit the criteria under consideration of evolving software technology and scientific standards, without requiring a completely new proposal being submitted. We envision funding periods of 5 years to provide sufficient security for funded software projects, while allowing for adaptation of the portfolio of funded software to novel research directions and community needs.
Failure to meet the reevaluation criteria should lead to the decision to phase-out sustainable funding. The phase-out process may come with a 1-year funding program based on a consolidation plan with clear goals regarding the archiving and preservation of the software, documentation, and all existing resources.

\section{Who Sustains Research Software?}
\label{sec:who_sustains}

Research relies on software and software relies on the people developing and maintaining it.
Sustainable research requires sustainable software, and this in turn requires continuity for those who develop and maintain it.

\subsection{Requirements}
Possibly the most important demand is the need for an increase in \textit{recognition and awareness} of research software as a first class citizen in research~\cite{Akhmerov2019,ChueHong2019,https://doi.org/10.2312/os.helmholtz.003}.
For sustainability of research software, long-term commitments of the respective software leads are crucial, but very few \textit{professional RSE profiles} currently exist.
In consequence, it is essential to create career paths for RSEs that are attractive and include permanency perspectives.
While creating permanent positions in the German academic system below the faculty level is an actively discussed topic overall~\cite{BayreutherErklaerung}, we specifically focus on the needs originating from the development and maintenance of research software here.

As already mentioned, research software development not only requires domain expertise, but also software development \textit{education, skills, and competencies}.
Currently, most of the domain researchers developing and maintaining domain-specific software technology never received professional training on software development~\cite{Wilson2014,SSIanalysis}.
To enhance the productivity and sustainability of computer-based research, it is essential to integrate software development training into the education of domain researchers.

Currently, a significant portion of the existing research software is developed by individuals or in small groups, primarily to serve their own requirements.
This situation is unsatisfying in terms of collaboration and inefficient in terms of several groups spending resources on generating similar or even the same functionality.
To enable and promote synergies, it is important to allocate resources for research software development and to build \textit{communities}, as described in~\cite{Katz19community}.

\noindent\vsepfbox{
  \parbox[]{\dimexpr\columnwidth-3\fboxsep-2\fboxrule\relax}{
    Kim wants to broaden her research portfolio within researchonomics and applies for postdoctoral positions at other institutions.
    Her library \emph{hal9k} is growing in popularity within researchonomics, and she wants to continue working on it.
    As her university has adopted an open science policy, \emph{hal9k} is free software under a Free/Libre Open Source Software (FLOSS) license, and Kim is free to continue her work on the library even after moving away from UofA.
    Due to her involvement in the creation of \emph{hal9k} as well as her previous success in attracting funding, Kim has the choice between multiple, attractive positions and decides to move to the researchonomics group at Eden University (EdU).
    She has already extended \emph{hal9k} in multiple directions in the past and  plans to continue this work at EdU.
    Her group leader at EdU would like to continue funding her but due to a law called the Fixed-term Research Contract Bill, EdU is not allowed to extend her contract, and neither third-party funding for her own position nor a permanent position are available.
    After having developed a now widely-used research tool, several publications in software and paper form, as well as having attracted funding, Kim finds herself looking for a job again.
  }
}

\subsection{Challenges}
We are currently facing a \textit{lack of awareness} for the importance of research software as discussed above. Moreover, there is \textit{little recognition} for the efforts put into software development and maintenance.
In consequence, software development in academic settings is mostly considered as a means to an end and sustainability is often not considered in project planning and grant proposals and contributes little to progressing research careers~\cite{Scienceguide,DORA}.
The main challenge here is the continued use of metrics that primarily leverage traditionally published articles and article citation numbers.

In academia, developers of research software are typically domain researchers, and in particular if new areas are explored, the software development process itself has research character.
Obviously, developing research software requires not only domain knowledge but also software development skills, and the researchers leading the software development process are often domain experts with substantial software development experience, making them extremely valuable members of the research community.
However, the current academic system in Germany does not provide a defined \textit{RSE role}.
Limited-term positions are, at least currently within the main German academic system, often effectively the end of their career path, sometimes even a dead end.
The challenge here is the lack of available permanent positions within the non-professorial academic faculty (``Mittelbau'') in Germany, compounded by a lack of access to 
these few permanent positions for RSEs due to the already mentioned lack of recognition for efforts concerning research software for faculty appointments within domain sciences.

In order to develop sustainable software, researchers need to have the \textit{skills and expertise} to build software that is easy to maintain and extend~\cite{carver2016software}.
However, most of the researchers are self-taught developers~\cite{Wilson2014,SSIanalysis}.
Ideally, these skills have to be built into the domain science curricula, which could generally be done in two different ways (or a combination of them).
One obvious solution attempt are additional courses that focus on these topics.
The main challenge here is to decide which other topic(s) to possibly drop due to the limited volume of any given curriculum.
A different approach is to incorporate software-related topics into existing domain science courses.
While this would provide the benefit of show-casing the usage of specific software skills directly within the domain science, the challenge here is the amount of work necessary to change existing lecture material, let alone the need of the lecturers to acquire those skills themselves in the first place.

As long as the necessary software skills within domain sciences are not yet wide-spread, building a network from those that have acquired relevant skills is difficult.
\textit{Community} efforts, that concentrate on questions regarding research software, can help to fill this gap.
Examples of such efforts include the Software Carpentries, national and international RSE societies (e.g., within Germany de-RSE e.V.).
However, since research software is such an interdisciplinary topic, it is hard to get recognition and find funding within any specific discipline.
As a result, existing communities often have to rely heavily on volunteers.
This is challenging, because despite benefits to domain science, volunteers hardly receive recognition for their work ``back home'', i.e., within their domain, underlining again the importance of our first demand.

\subsection{Recommendations}
Increasing \textit{recognition and awareness} is a challenge that calls for both immediate action and perseverance. Nevertheless, some measures will likely show positive effects comparatively soon.

Similarly to plans for research data management, funding agencies should request that applicants include considerations about how software developed in a project can be sustained beyond the end of the funded project. A follow up on these plans during and after the project lifetime, i.e., a dedicated software management plan, is crucial.

Another recommendation is aimed at decision makers concerning recruitment for academic positions: broaden the definition of research impact beyond traditional scientific publications to also include other impactful results.
Not all researchers thinking of themselves as RSEs pursue a faculty position as their main career goal.
However, permanent academic non-faculty positions are rare within the German academic system, also due to the lack of a defined \textit{RSE role}.
We recommend research institutions to leverage the benefit of dedicated RSEs by establishing attractive long-term career options in the academic environment.
The long-term solution in order to gain sufficient software development \textit{skills} should be education that is included early in the career path, ideally already at the Bachelor level.
For the time being however, efforts involving workshops and seminars that provide easy access to hands-on training on software-related questions should be promoted and supported as much as possible.

It is important to provide an environment where \textit{communities} can form and flourish by allocating resources for research software development and for building communities around it~\cite{Katz19community,iaffaldano2019developers,Jimnez2017}.
The identification with a community of like-minded people and personal action~\cite{dagstuhl} can lead to a permanent establishment of sustainable research software as a valuable research output.
Thus, research institutions as well as funding agencies should not only be open-minded regarding existing volunteer organizations, but should actively promote the creation of such groups.

\section{How can Research Software be Sustainably Funded?}
\label{sec:funding}

\noindent\vsepfbox{
  \parbox[]{\dimexpr\columnwidth-3\fboxsep-2\fboxrule\relax}{
    \emph{Hal9k} has grown into a widely used software in researchonomics, and Kim is proactively asked to apply for - and is subsequently awarded - a permanent RSE position at the institute for researchonomy at UofA, based on her work on the library. She works closely with the central RSE team, but mostly due to bureaucracy and the high demand for her library, Kim does not have enough time to maintain and further develop \emph{hal9k} alone anymore. Together with the dean she develops a course for the researchonomics curriculum which teaches data processing with \emph{hal9k}. As a lesson from her own career, she starts the course with sessions on the Hash shell, version control with Tig, Boa, and two whole sessions on basics of sustainable software development. This is very fruitful, and due to the implementation of a new research software funding scheme at UofA, Kim is able to hire one of the course students, who has shown great RSE skills, straight into a long-term position at her institute, where they focus on the maintenance and development of \emph{hal9k}, work with the computing center to support \emph{hal9k}-based supercomputing on a new, dedicated FGPA cluster, develop training materials for external users, and organize the yearly \emph{hal9k} users and developers conference. Kim gets to travel the world to visit researchonomics groups who are using \emph{hal9k}.
  }
}

\subsection{Requirements}
Sustainable funding for research software boils down to funding the four main pillars enabling sustainable software development: 
\begin{inlinelist}
    \item Personnel with expertise in research software development;
    \item Infrastructure for developing, testing, validating, and benchmarking research software, and distributed versioning systems for collaborative software development;
    \item Training in software design and sustainable software development; and 
    \item Community management and events for creating synergies between research groups and software efforts.
\end{inlinelist}

\subsection{Challenges}
Short-term engagement of (young) researchers raises the question of how to maintain a
constant level of expertise within a developer team and prevent knowledge drain concerning domain knowledge and software engineering skills.
Conversely, the permanent engagement of qualified personnel requires to offer career perspectives, especially due to the fact that academia competes with
industry for the same people. A challenge specific to Germany is posed by the
shortage of permanent positions and by the restrictions for temporary positions
due to the German \emph{Wissenschaftszeitvertragsgesetz}~\cite{BMJVarbeit}.

Sustainable software development requires hardware technology to develop, test, validate, and benchmark features in a continuous integration cycle. The challenge in this context is the persistent evolution of the hardware landscape. Hence, for creating an environment promoting sustainable software development, it is important to provide access to a wide hardware portfolio and to support a development cycle based on continuous integration.

Expertise in sustainable research software development is a scarce resource, and training is heavily needed as one way of building up more expertise. However, while integrating interdisciplinary software engineering courses into the education curriculum can build up basic skills, some expertise is domain-specific and requires inter-institutional training activities. Furthermore, there exist no financial incentives for creating software-specific documentation and tutorials nor to provide other forms of support.

While the creation of research software communities is one of the major assets in sustaining research software technology, promoting this process requires the installation of new funding instruments. Traditionally, research grants are limited to rather short time frames and support personnel, material, hardware, and to a limited degree also travel and research visits. Creating a research software community however requires funding for community and training events as well as ``virtual hardware'' such as webspace, versioning systems, task-managing systems, and compute cycles. These demands can hardly be met without third-party funding~\cite{Kuchinke2016,chang2007open,Aartsen2018,Gabella2018}.

\subsection{Recommendation: Creation of Adequate Funding Schemes}

Funding is a crucial factor for sustaining research software. Currently available sources and instruments are not adequately shaped for the challenges and solutions outlined above. We recommend actions on the individual, organizational, and national level.

Existing project-focused funding instruments on the local, national, and international level need to be complemented with funding instruments specifically designed for research software development and sustained research software maintenance to make research software a first class citizen in the research landscape. For example, software projects enhancing research and fulfilling the sustainability criteria detailed in section~\nameref{sec:selection} on p.~\pageref{sec:selection} may be entitled for sustained funding as long as they live up to the standards and remain a central component of the research landscape.

Computing centers and supercomputing facilities for research need to receive earmarked resources for the support of sustainable software development. This funding is necessary to provide continuous integration services, a hardware portfolio for development, testing and benchmarking software, as well as personnel for training domain researchers in software design and the proper usage of the services.

The creation and maintenance of training materials for general research software engineering education and the software-specific documentation and tutorial creation needs to be reflected in funding opportunities. This can either happen by dedicating modules of research or software grants to providing support and the generation of training material, or by opening funding schemes focusing on interdisciplinary software development education. The latter may include research that looks at research software development as a process to analyze which measures, interactions, and team compositions make research software successful. Additionally, funding instruments fostering the formation of research software communities have to be established.
\vfill\null 

\section{Which Infrastructure is Needed to Sustain Research Software?}
\label{sec:infrastructure}
\noindent\vsepfbox{
  \parbox[]{\dimexpr\columnwidth-3\fboxsep-2\fboxrule\relax}{
    As the \emph{hal9k} community grows, so does the need for infrastructure. Kim and her team collaborate with the National RSE Consortium to set up \emph{hal9k} on the Consortium’s distributed TigHub instance, and organize world-wide access to it via the NRSEC-AAI federation. Going forward, the Consortium’s Research Software Hub - a registry and Software Heritage Archive-based~\cite{softwareheritage} long-term repository for research software on a national level - ingests \emph{hal9k} releases with complete metadata: citation metadata, the \emph{hal9k} provenance graph and computational environment information, ORCID iDs~\cite{ORCiD}, etc. and provides its own DOIs for versions under a concept (umbrella) DOI.
    The community reviews all code and documentation changes that are contributed to \emph{hal9k} via the central TigHub, and the Hub’s CI system Alfred builds, tests, and pushes new releases automatically to the registered supercomputing clusters.
    Especially the community efforts become better and more streamlined by the day, as research software development training is now offered as part of most curricula, and skilled RSEs are now much easier to find and hire by research institutions.
  }
}

\subsection{Project Management Tools}
\label{sec:infrastructure:PM}
Research software is developed by individual researchers, in small teams within a single institution, or in larger teams distributed across multiple institutions. In particular if software development is distributed across institutions, there exists an urgent need for frameworks and tools enabling collaborative code development, software feature planning, and software management. As research software development typically includes bleeding-edge research and development that the researchers do not want to disclose for a certain time to preserve intellectual property, distributed research software development also needs a global Authentication and Authorization Infrastructure (AAI). We recommend the development and/or deployment of tools for distributed software development and software management as central research infrastructure. An important aspect in this context is the cataloging of research software to reduce the duplication of development efforts. This can efficiently be realized by promoting the registration of all research software with a unique identifier and developing a tool that allows to explore the research software landscape. Research software contributors should have an ORCID iD~\cite{ORCiD} to be uniquely identifiable and referable. While some funding for such tools and software repositories is emerging (e.g. the bio.tools catalogue of bioinformatics tools funded as part of the European ELIXIR project~\cite{bio.tools}), a standardized extension of such efforts to the RSE community as a whole is necessary. However, as the experiences from ELIXIR
demonstrate, this is a non-trivial effort that requires significant dedicated
and long-term funding.

\subsection{Developer Training, Motivation, and Knowledge Exchange}
As elaborated, training in sustainable software development is key to achieve sustainability in research software. At the same time, it is not clear how such training should be facilitated and institutionalized. Furthermore, for deriving software quality standards, evaluating the quality of software, and providing a code review service, central resources are necessary that individuals and groups in the research software landscape can draw from.

We consider Software Carpentry and similar efforts like the creation of the Data Science Academy HIDA~\cite{HIDA} in the Helmholtz Association of German Research Centers helpful solutions to exchange and distribute knowledge. Local chapters of RSE groups and (inter-)national conferences will further foster networking and community building. We strongly recommend the creation of a national Software Sustainability Institute (involving funded positions to establish web platforms and training material) similar to the existing institute in the UK~\cite{UKSoftwareSustainabilityInstitute}, which serves as a national contact for all aspects related to research software. The UK SSI also publishes best practice guidelines~\cite{SoftwareSystemsDevelopmentLifeCycle_SDLC} for research software engineering.

\subsection{Research Software Discovery and Publication}
Proper software publication and possibilities for the community to find existing software solutions for a given problem are a prerequisite to optimally exploit synergies and avoid redundant development. However, we observe that today, many funding proposals lack a thorough state-of-the-art report of software that could possibly be reused. This is most often caused by insufficient information retrieval strategies, lack of knowledge about relevant repositories, and an abundance of locations where software is collaboratively developed and stored~\cite{Struck2018}. Discovery requires publication in a globally accessible location with appropriate metadata, e.g., Citation File Format (CFF)~\cite{CFF} and CodeMeta~\cite{CodeMeta}. Comprehensive metadata (e.g. contributors, contact, keywords, linked publications, etc.) and publishing platforms have to enable persistent citing, which in turn benefits research evaluation. Selection and curation of software (probably by a data/software librarian) for publication and discovery are certainly challenging.

We consider GitLab or GitHub as collaborative working environments and repositories like Zenodo appropriate publication platforms, because the latter mint DOIs, allow versioning and are publicly funded for long-term access. GitHub, Figshare, and Mendeley Data are examples of commercial enterprises with business cases in the background, which leverage research results. Besides the aforementioned metadata standards, it is advisable to document source code, e.g., using MarkDown (with Doxygen tooling). Metadata and citations play a role in beneficial tools like PIDgraph, DataCite.org, CrossRef, which utilize Persistent Identifiers (PIDs) like DOIs.
Another solution to discovery are (mostly) disciplinary software indices like swMATH~\cite{swMATH} or the Astronomy Source Code Library~\cite{ASCLnet} as well as language focused systems like CRAN~\cite{CRAN} for \textsf{R}. Most of them started as national endeavors and became platforms of global importance. For Germany, we assume that the Nationale Forschungsdateninfrastruktur (NFDI) will put effort into creating or supporting discovery platforms at a central point that ease information retrieval. At the same time, all stakeholders should be aware of and counteract potential institutional ``fear'' of losing ``their'' data, software, and intellectual property.

Especially in interdisciplinary environments, it would be helpful to have access to a meta software repository index, similar to what re3data~\cite{re3data} does for research data repositories. We recommend the creation of such a meta index covering important (disciplinary) software indexes in order to ease discovery of relevant software locations. Evaluation of discovered software is an unsolved problem. Here, anonymous telemetry of usage may provide information for the selection of relevant software. Publishing software, their dependencies, and environment in containers may also ease evaluation and further reuse. These suggestions require significant investment in long-term infrastructure.
When publishing research software it is recommended to make use of integration schemes like GitHub with Zenodo or local GitLab instances with publication platforms.
Such indices and publication outlets may benefit national federated research indexing \& archiving systems, similar to the hierarchy of library catalogs~\cite{KVK}.

\subsection{Archiving}
Software preservation aims to extend the lifetime of software that is no longer actively maintained. There are different approaches, which vary in the effort required and the likelihood of success. Software archiving is one important aspect of software preservation: the process of storing a copy of a software so that it may be referred to in the future.
The publication of a certain software version for reference in research articles requires simple ways to archive research software on a long-term basis. Furthermore, its integration with collaborative software development environments such as GitLab or GitHub and with publication repositories is needed to facilitate archiving of referenced software versions based on sustainable frameworks (e.g. Invenio~\cite{Invenio} for GitHub to Zenodo integration).

A challenge for software archiving is the need to (ideally) preserve the runtime environment and all dependencies of the software. This could improve reproducibility, especially when running the software in its original state. If research data are needed to reproduce results, they should also be archived with the software or the publication. Specialized and unique hardware - like high performance computing resources - can be part of the runtime environment, which may not be accessible in the future. To overcome this, an emulation of hardware may be a (challenging) solution. Emulation involves the encapsulation and distribution of the complete hardware and software stacks, including the operating system and driver interdependencies. This can result in intellectual property issues when offered as a service.

There are both local and global approaches to software conservation. One solution to keep the software in an executable state by preserving its context and runtime environment is to use containers such as Docker.
However, to archive the Docker containers, additional metadata should be added and stored with the software in an archive container format that allows exchange between repositories and exit strategies, such as the BagIt container format~\cite{kunze2018bagit}. Application or platform conservation is also achieved by conservational efforts where unmaintainable (virtual) machines are sandboxed to keep the platform in a secure but running state.
Another threat is losing project repositories on global platforms like Github or BitBucket. Here, global platforms like Software Heritage~\cite{swHeritage} harvest those repositories and prevent loss by long-term archiving.

\section{Legal Aspects}

\noindent\vsepfbox{
  \parbox[]{\dimexpr\columnwidth-3\fboxsep-2\fboxrule\relax}{
    More and more industrial partners enter the \emph{hal9k} community, and they bring their lawyers. Together with UofA’s research software task force, the RSE team, the researchonomy institute, the corporate lawyers, and community representatives, Kim decides to create a foundation to govern \emph{hal9k} and its environment: the Fullest Possible Use Foundation for Open Researchonomy, funded by the Ministry of Research and Education and a consortium of corporate partners. As a first step, they re-license \emph{hal9k} under the OSI approved MIT license.
  }
}


\noindent A common situation in research software creation is that the developer has no knowledge or awareness of legal aspects and therefore did not consider them early enough. Thus, we think the main legal demands for research software development are raising awareness and empowering all levels of responsible persons in academia (from researcher and RSEs over PIs to research performing organizations and research funding organizations) in legal aspects. This will hopefully lead to a general legal certainty before, during, and after the research software development process and thus enable better options for collaborations between universities, non-commercial research institutions, and other national or international partners. Legal aspects always have to be considered regarding the relevant jurisdiction. Though similar issues arise in all jurisdictions, the following will focus on the European and specifically German legal framework.

\subsection{Challenges and Clarifications}
\subsubsection{Clarification of Rights}
Software development is a creative activity. The main relevant law governing legal aspects is therefore the copyright law. It regulates the rights and obligations of the parties involved. Chapter 8 of the German Act on Copyright and Related Rights (UrhG) contains specific provisions applicable to computer programs and is based on the EU computer programs directive. Copyright law protecting the creator of software in similar ways exist in nearly all legal systems. It is important for the identification of rights that software, in the sense of (German) law, includes not only the source code but also the design materials~\cite{BMJVurheber69a}. The challenge in the use, distribution, and commercialization of software is to determine the chain of rights and to identify all right holders. The owner of the copyright is not necessarily the owner of the right of use. For Germany, the Copyright Act regulates the rights for employment relationships~\cite{BMJVurheber69b}. In such cases, the right of use is automatically transferred to the employer. This means that in most cases of employed software developers and research staff, the institution holds the rights of use for the software work. This is not automatically the case for students, freelancers, and individual external cooperation partners. Employment and service contracts with contributors could contain regulations regarding the transfer of rights of use. For researchers who conduct free research not subject to directives, in Germany the constitution guarantees freedom of research so that the rights of use for their work remains initially with the natural person. In addition to the rights of the people directly involved, other rights of third parties may also be relevant. Existing source code (e.g., other Free/Libre Open Source Software (FLOSS)), external libraries, and contributions from institutional cooperation partners are published and provided under certain licenses and their conditions must be observed (which, due to incompatibilities even among FLOSS licenses,
may well mean that individually reusable pieces of software cannot be 
reused together or in a new context). The nature of research careers often brings additional complications to the chain of rights. It happens that researchers take their software with them when they change institutions and develop it further during their career. Here, the former employer may be entitled to some rights of use. In third-party funded projects, in particular with industry but also with public funding, rules regarding rights of use are often defined. Last but not least, the software can also be affected by other (intellectual) property rights such as patents or trademarks. Software itself is usually not patentable but it may implement a technical invention covered by patents. When using or distributing such software, an additional matching patent license may be necessary. Licenses exist (for example: GNU GPL v3) which automatically grant related patent licenses while using the software license. That should be considered when exploitation of the patent is planned.

\subsubsection{Liability}
Issues of warranty and liability for faulty software must be taken into account. We consider the possibilities of contractual limitation of liability in licenses. Full exclusions of liability are generally invalid in the German law. Limitations of liability usually depend on the form of distribution: The limitation options are larger if the rights of use are granted free of charge, e.g. provision ``as is'' as defined in the BSD 3-clause license.

\subsection{Ideas for Solutions}
In order to meet the legal challenges mentioned, it is absolutely necessary for the software developer (team) to document the rights chain comprehensively during the software development (see, e.g., supplementary material). Contributions of individual persons must be traceable and their (labor law) status must be named. At best, contracts with rules on the transfer of rights of use should be concluded before work begins. Declarations of assignment of rights can be made for existing works. License conditions for external contributions must be evaluated with regard to further rights of use and possible sub-licensing. Contracts and funding conditions must be conscientiously documented and analyzed with regard to rules on rights of use. In case that different parts of the software are based on different conditions and rights of third parties, individual modules of the new software could be published under different licenses and merged accordingly.

A national research software sustainability institute could be established. This institute supports local research software task forces and thereby respective researchers and research teams in the licensing of research software and related legal issues. For this purpose, a legal help desk will be set up, to which all members of their respective research performing organization can apply. If researchers want to publish the research software under a Free/Libre Open Source Software license, the organization could bundle the necessary rights beforehand. This is particularly useful when teams of researchers, often international, write software. In addition, the sustainability institute may serve as a one-stop-shop for the licensing of research software.

\subsection{Recommendations}
We see it as an essential part of the sustainability of research to enable the free distribution of research software. There are a variety of open source software licensing models (ranging from permissive to copyleft; for further information, see~\cite{ifrOSS,tldrlegal,morinQuickGuideSoftware2012}). The use of an FSF- or OSI-approved FLOSS license for example would enable a truly free model and also reduce legal issues. We recommend that research funding organizations such as the DFG discuss if they expect publishing all funded software under these licenses, following the paradigm of ``public money, public code''~\cite{publiccode}.

\begin{figure}[tb]
  \centering
  \includegraphics[width=\linewidth]{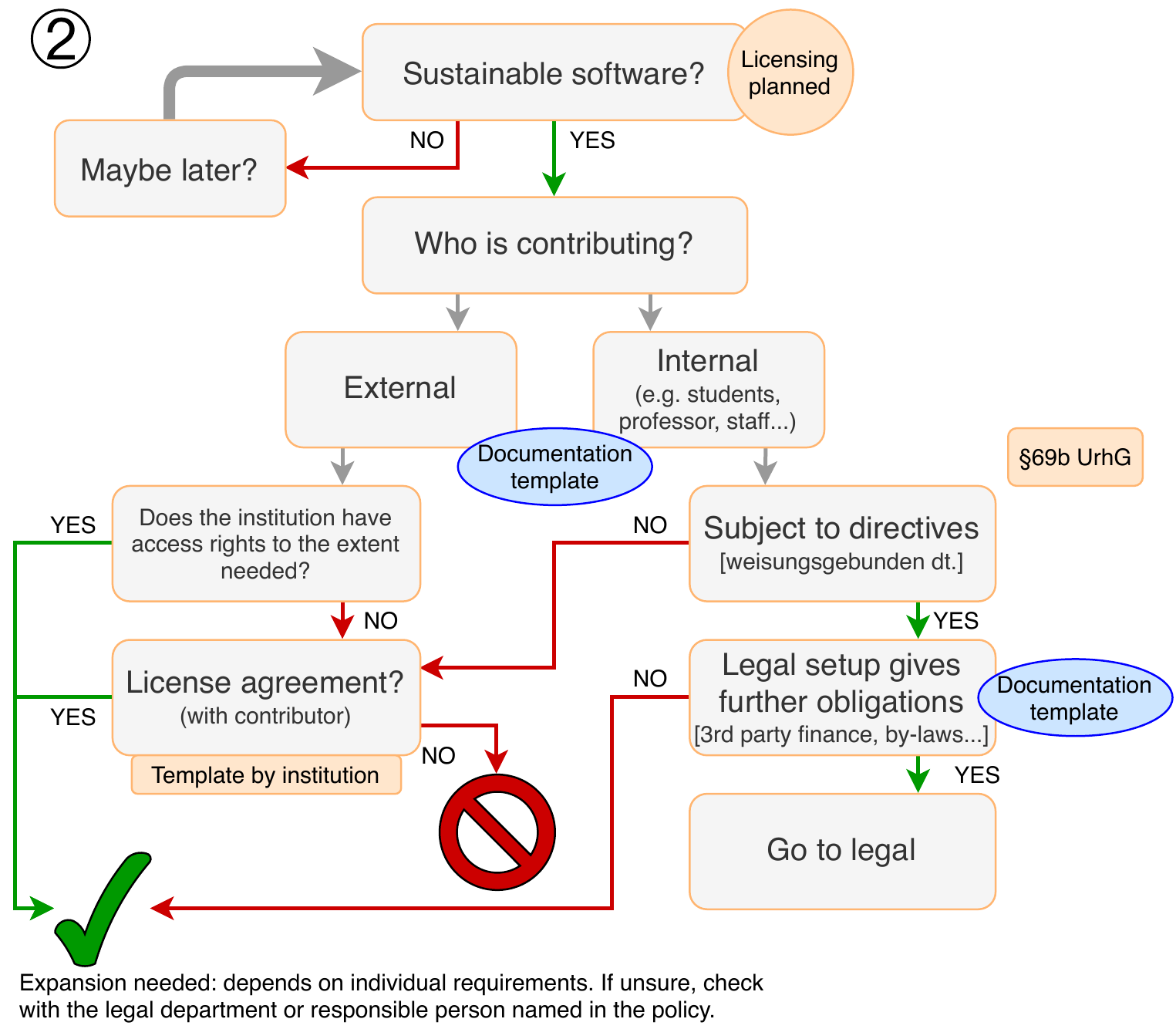}
  \caption{\textbf{Decision tree for contributors}. This tree helps to figure out whether the academic institution where the software is developed owns the intellectual property (copyright).}
  \label{fig:tree1Article}
\end{figure}

Also for legal aspects, we believe it is important that all (German) research performing organizations install a research software task force, especially since the new DFG Code of Conduct~\cite{DFGcode} was released. Besides organization and bundling of technical and infrastructural support for local RSEs and researchers (see previous sections), this group should organize a local legal help desk, organize educational offers e.g. for the legal topics presented, and (if not implemented yet) develop the software policy of the research performing organization. As an example, with the help of on-boarding processes performed by the research software task force, RSEs should be able to keep the clearance of rights as simple as possible right from the start. One possibility how local legal help desks could structure their work is shown in the decision tree in Fig.~\ref{fig:tree1Article}. A more complete suggestion for decision trees for both legal help desks and interested RSEs can be found in the supplementary material. We suggest that the local task forces build a network with the other research performing organizations for exchange of ideas but also for generating a bottom-up strategy to organize RSE standards for Germany and beyond and possibly be the origin of the aforementioned software sustainability institute.

\section{Conclusions}
We find that the research software ecosystem is notoriously lacking resources despite its strategic importance. If funding and support does not improve, the success story of science based on academic research software may be at stake. 
We recommend the installation of infrastructure that enables sustainable software development including platforms for collaboration, continuous integration, testing, discovery, and long-term preservation.
We suggest the establishment of a nationwide institution similar to the Software Sustainability Institute (SSI) to provide project consulting and code review services as well as sustainable software development training. We think that sustainable software development should become an integral component of the universities' teaching curriculum.
We encourage the research funding bodies to reflect the licensing models for academic software development, and to decide whether the ``public money, public code'' paradigm justifies the requirement that all publicly-funded software has to be publicly available under a Free/Libre Open Source Software (FLOSS) license.
Ultimately, we strongly advise the implementation of funding schemes for sustainably supporting the development and maintenance of research software based on clear and transparent criteria, for creating incentives to produce high quality community software, and for enabling career paths as research software engineer (RSE).

\section{Declarations}

\subsection{Glossary}\label{subsec:glossary}

\printnoidxglossaries









\subsection{Consent for Publication}

Not applicable



\subsection{Competing Interests}
The authors declare that they have no competing interests.


\subsection{Funding}
The authors thank the DFG for funding a meeting (\textit{Rundgespr\"{a}ch}, grants LO
2093/3-1 and SE 1758/6-1) 
during which the initial draft of this paper has been created. We are particularly grateful for the support from Dr.\ Matthias Katerbow (DFG).

This work was additionally supported by Research Software Sustainability grants funded by the DFG~\cite{DFGfunded}:
    Aggarwal: 390886566; PI: Zesch.
    Appel: 391099391; PI: Balmann.
    Bach \& Loewe \& Seemann: 391128822; PIs: Loewe / Scholze / Seemann / Selzer / Streit / Upmeier.
    Bader: 391134334; PIs: Bader / Gabriel / Frank.
    Brusch: 391070520; PI: Brusch.
    Druskat \& Gast: 391160252; PI: Gast / L\"udeling.
    Ebert: 391137747; PI: Marschall.
    Flemisch \& Hermann: 391049448; PIs: Boehringer / Flemisch / Hermann.
    Hohmann: 391054082; PI: Hepp.
    Goth: 390966303; PI: Assaad.
    Grad \& Weeber: 391126171; PI: Holm.
    Kutra: 391125810; PI: Kreshuk.
    Mehl \& Uekermann: 391150578; PIs: Bungartz / Mehl / Uekermann.  
    Peters-Kottig: 391087700; PIs: Gleixner / Peters-Kottig / Shinano / Sperber.
    Raters: 39099699; PI: Herwartz.
    Reina: 391302154; PIs: Ertl / Reina.
    Muth \& Renard: 391179955; PIs Renard / Fuchs.
    Ropinski: 391107954; PI: Ropinski.


\subsection{Author's Contributions}
We are a group of software-providing researchers, RSEs, and infrastructural as well as legal supporters. Initially, a group of representatives of funded projects of the first DFG sustainability call~\cite{DFGfunded} met during the first German RSE conference (deRSE19)~\cite{deRSEconf} in June 2019 in a grass-roots workshop on sustainable research software addressing the software-based research community. During this workshop, we realized that a lot of valuable experience and good ideas are present in the group, and we decided to start working on this paper together with other interested practitioners. We followed the generous invitation of the DFG for the above-mentioned two-day meeting at the Robert Koch Institute in Berlin in November 2019 to sharpen the focus of this paper.\\
The individual contributions of authors to this work are detailed below, following the \href{https://casrai.org/credit/}{CASRAI CRediT} (Contributor Roles Taxonomy):
\begin{itemize}
    \item \textbf{Conceptualization:}  Anzt, Bach, Druskat, Loewe, Löffler,
    Renard, Seemann, Struck
    \item \textbf{Funding acquisition:} Loewe, Seemann
    \item \textbf{Investigation:} Bach, Druskat, Loewe, Löffler, Renard,
    Seemann
    \item \textbf{Project administration:} Loewe, Seemann
    \item \textbf{Visualization:} Unger, Friedl, Löffler
    \item \textbf{Writing original draft:} Achhammer, Aggarwal, Anzt, Appel, Bach, Bader, Brusch, Druskat, Ebert, Flemisch, Friedl, Funk, Grad, Goth, Herrmann, Hohmann, Kutra, Linxweiler, Loewe,
    Löffler, Muth, Peters-Kottig, Rack, Raters, Rave, Reina, Renard, Ropinksi, Schaarschmidt, Seemann, Struck, Thiele, Uekermann, Unger, Weeber
    \item \textbf{Writing review \& editing:} Anzt, Appel, Bach, Bader, Brusch, 
    Busse, Chourdakis, Dabrowski, Druskat, Friedl, Fritzsch, Funk, Gast, Herrmann, Janosch, Loewe, Löffler, Rack, Reina, Reißig,
    Renard, Seemann, Seibold, Struck, Thiele, Uekermann, 
\end{itemize}




\bibliography{references}

\clearpage

\renewcommand{\thefigure}{S\arabic{figure}}
\setcounter{figure}{0}  
\section{Supplementary Material}
\subsubsection{Decision Trees for Legal Topics}
The decision trees presented here shall help legal help desks and developers to identify risks regarding the mandate of the software. In a perfect world, one would address the legal aspects at the start of a project. It is crucial to know about these to create sustainable software. We strongly recommend to write a documentation of the answers and outcomes. Please keep in mind that only restrictions from copyright law are addressed. In some projects, you might also have to consider patents, trademarks etc.

\begin{figure}[h!]
  \centering
  \includegraphics[scale=0.5]{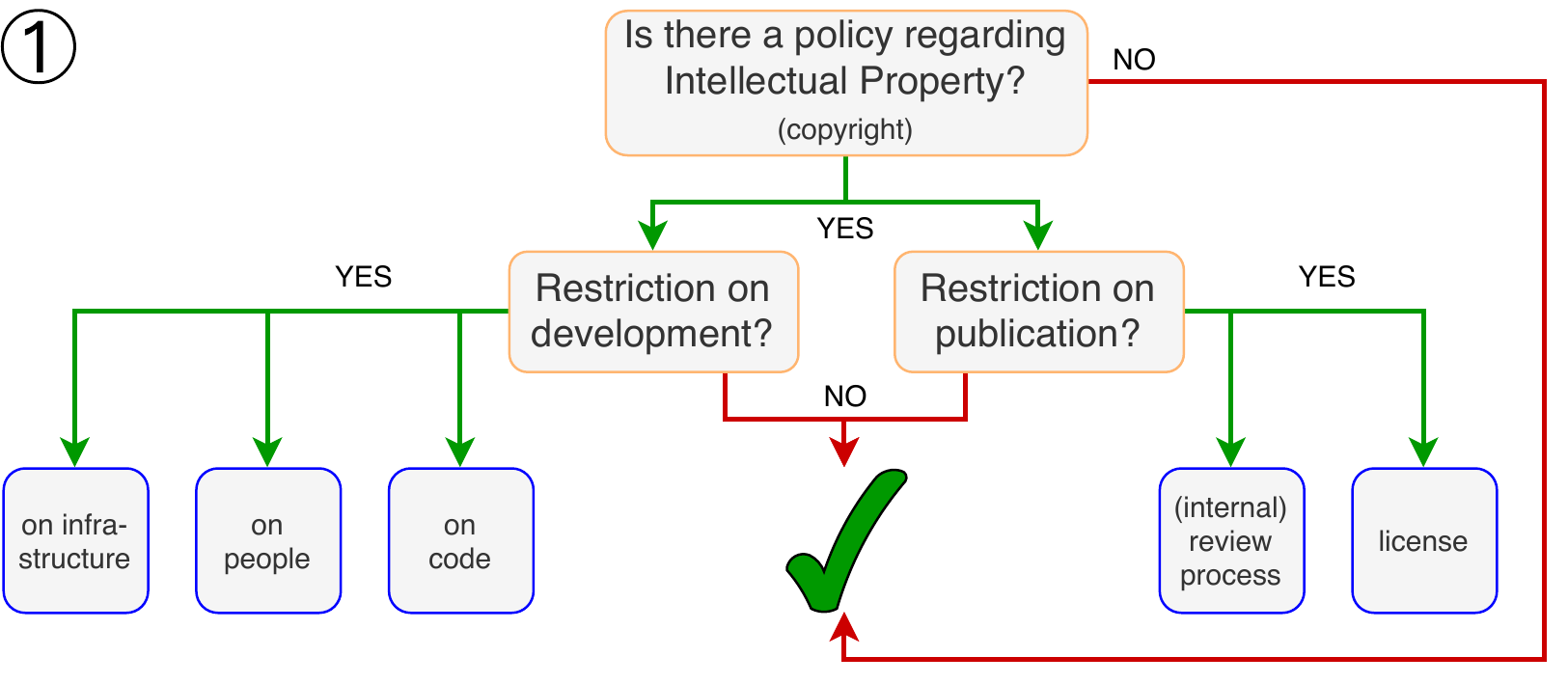}
  \caption{\textbf{Policy}. This tree recommends to check closely any policies implemented in the software developers organization.}
  \label{fig:tree0}
\end{figure}

\begin{figure}[h!]
  \centering
  \includegraphics[scale=0.5]{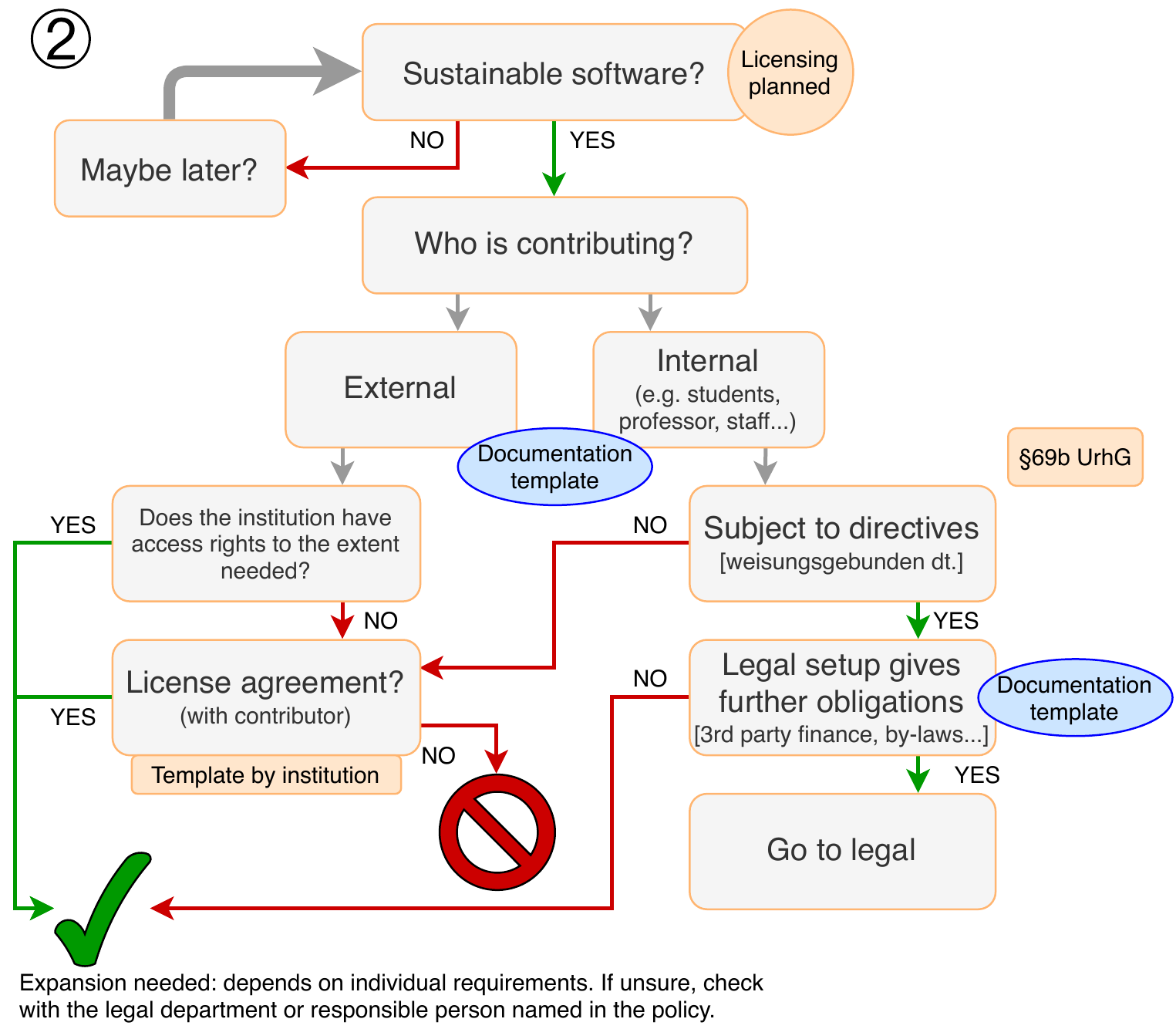}
  \caption{\textbf{Contributors}. This tree helps to find out whether the academic institution where the software development is located is the owner of the intellectual property (copyright).}
  \label{fig:tree1}
\end{figure}

\begin{figure}[h!]
  \centering
  \includegraphics[scale=0.5]{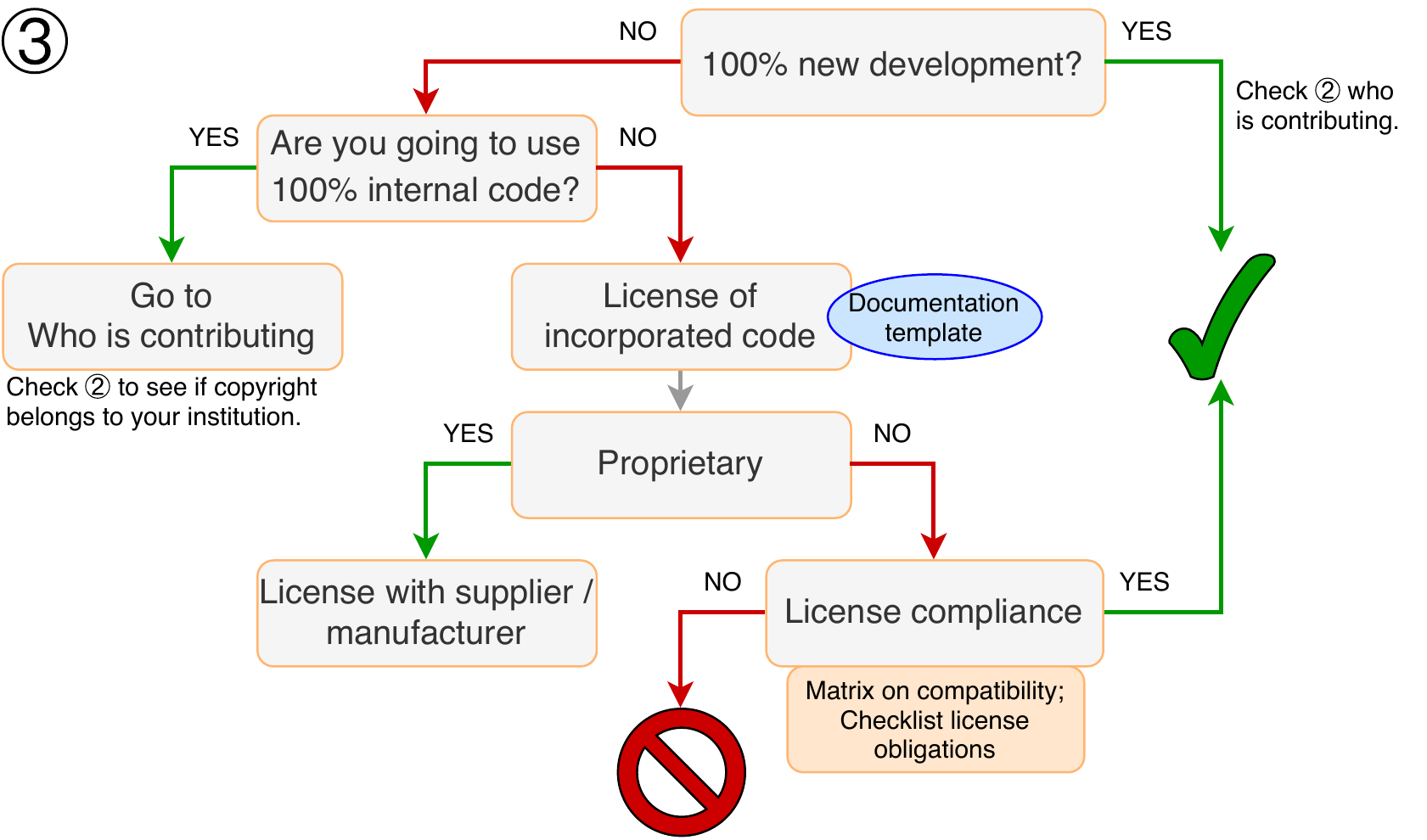}
  \caption{\textbf{Code history}. The code history tree points out tasks for projects that incorporate existing code.}
  \label{fig:tree2}
\end{figure}

Before you can publish, use, and/or license a software, you have to check:
\begin{itemize}
    \item The policy of the institution (Fig.~\ref{fig:tree0})
    \item The rights restriction imposed by the persons who ``create'' the software (Fig.~\ref{fig:tree1})
    \item The rights restriction imposed by the environment (Fig.~\ref{fig:tree1})
    \item If third-party code is incorporated (Fig.~\ref{fig:tree2})
\end{itemize}

\begin{figure}[h!]
  \centering
  \includegraphics[scale=0.5]{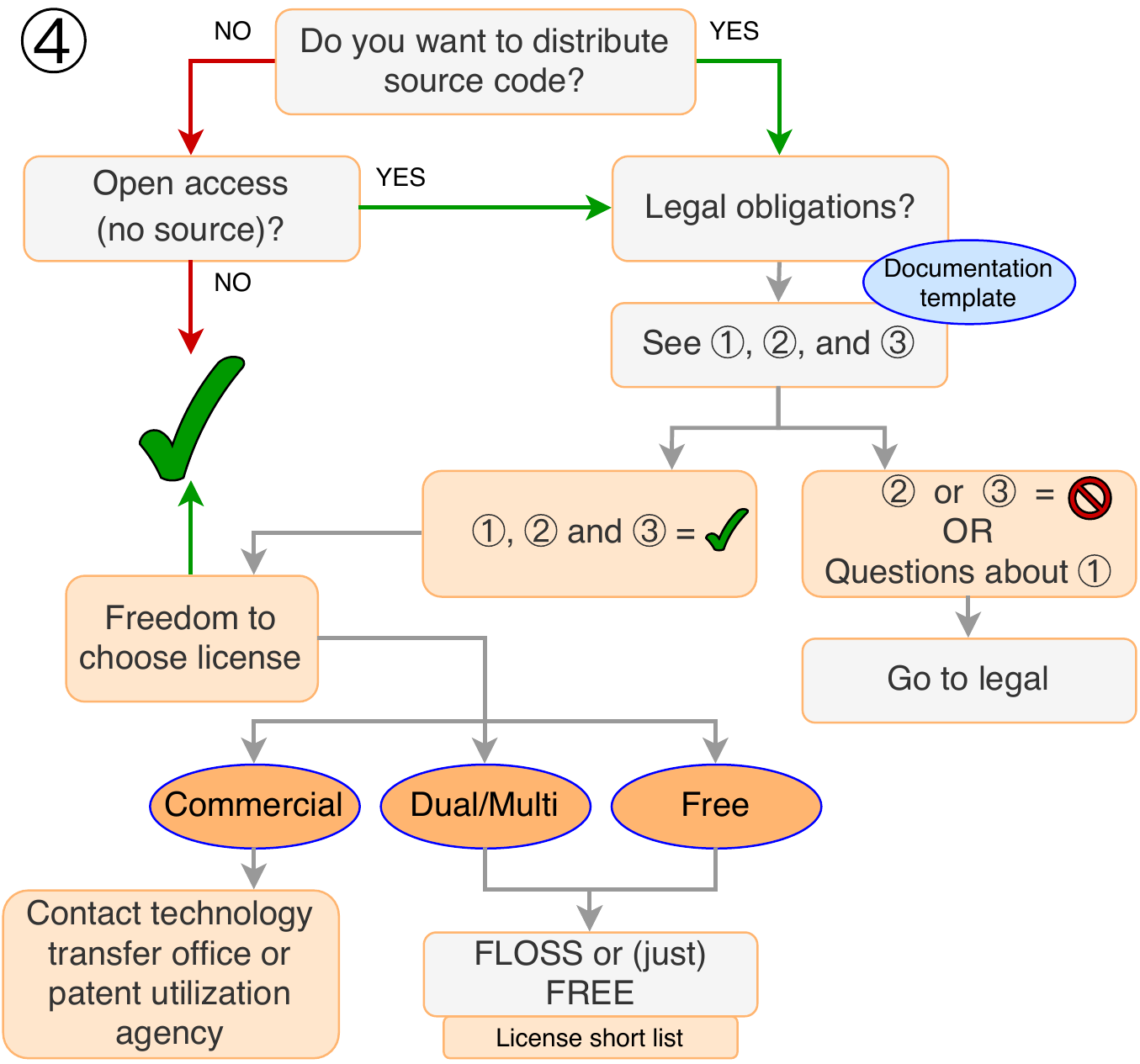}
  \caption{\textbf{Licensing}. Depending on the distribution model, open access (OA) or open source software can be selected.}
  \label{fig:tree3}
\end{figure}

We also built a tree for the scenario that you have to check for already existing software (Fig.~\ref{fig:tree3}).

If the outcome is a prohibition sign, we believe there is no other solution than to rewrite parts of the code or the whole code. If the outcome is a green checkmark, we believe you have the rights which you need to proceed. The other outcomes are self-explaining (e.g. go to legal department).

\end{document}